\documentclass[fleqn,usenatbib]{mnras}
\usepackage{amsmath}
\usepackage{xcolor, graphicx}
\usepackage{amssymb}
\usepackage{soul}
\usepackage{txfonts}
\usepackage{caption}

\begin{document}

\newcommand{\re}{R_{\rm e}}
\newcommand{\Msun}{\rm M_\odot}

\title[Hot gas total mass connection]{The SLUGGS Survey: revisiting the 
correlation between X-ray luminosity and 
 total mass of massive early-type galaxies}
\author[D. A. Forbes et al.]{Duncan A. Forbes$^{1}$\thanks{E-mail:
dforbes@swin.edu.au}, Adebusola Alabi$^{1}$, Aaron J. Romanowsky$^{2,3}$, Dong-Woo Kim$^{4}$, 
%Jay Strader$^{4}$
\newauthor 
Jean P. Brodie$^{3}$ and Giuseppina Fabbiano$^{4}$
%Christopher Usher$^{5}$, Vincenzo Pota$^{6}$\\
%Jay Strader$^{3}$\\
\\
$^{1}$Centre for Astrophysics \& Supercomputing, Swinburne
University, Hawthorn VIC 3122, Australia\\
$^{2}$Department of Physics and Astronomy, San Jos\'e State
University, One Washington Square, San Jose, CA 95192, USA\\
$^{3}$University of California Observatories, 1156 High Street, Santa Cruz, CA 95064, USA\\
$^{4}$Smithsonian Astrophysical Observatory, 60 Garden Street, Cambridge, MA 02138, USA\\
%Department of Physics and Astronomy, Michigan State
%University, East Lansing, Michigan 48824, USA\\
%$^{3}$Department of Physics and Astronomy, Faculty of Sciences,
%Macquarie University, Sydney, NSW 2109, Australia\\
%$^{4}$Australian Astronomical Observatory, PO Box 296 Epping, NSW
%1710, Australia\\
%$^{5}$Astrophysics Research Institute, Liverpool John Moores University, 146 Brownlow Hill, Liverpool L3 5RF, UK\\
%$^{6}$INAF - Osservatorio Astronomico di Capodimonte, Salita Moiariello, 16, 80131 Napoli, Italy
}

%\date{Accepted 1988 December 15. Received 1988 December 14; in original form 1988 October 11}

\pagerange{\pageref{firstpage}--\pageref{lastpage}} \pubyear{2002}

\maketitle

\label{firstpage}

\begin{abstract}

%The scatter in X-ray scaling relations, the heating mechanisms, and even
%the origin, of the hot gas in the halos of early-type galaxies continue
%to be subjects of debate. 
Here we utilise recent measures of galaxy
total dynamical mass and X-ray gas luminosities (L$_{X,Gas}$) for a sample
of 29 massive early-type galaxies from the SLUGGS survey to probe
L$_{X,Gas}$--mass scaling relations. In particular, we investigate
scalings with stellar mass, dynamical mass within 5 effective radii
(R$_e$) and total virial mass. We also compare these relations with
predictions from $\Lambda$CDM simulations. We find a strong linear
relationship between L$_{X,Gas}$ and galaxy dynamical mass within 5
R$_e$, which is consistent with the recent cosmological simulations of Choi et al. that
incorporate mechanical heating from AGN. 
We conclude that the gas
surrounding massive early-type galaxies was shock heated as it fell
into collapsing dark matter halos so that L$_{X,Gas}$ is primarily
driven by the depth of a galaxy's potential well. Heating by an AGN
plays an important secondary role in determining L$_{X,Gas}$.

%The $\sim$100-1000$\times$ 
%scatter seen in the 
%L$_{X,Gas}$--stellar mass relation is largely a consequence of 
%the halo mass-to-stellar mass relation. 

\end{abstract}

\begin{keywords}
galaxies: star clusters -- galaxies: evolution -- galaxies: Xrays
\end{keywords}

\section{Introduction}

Massive early-type galaxies (ETGs) reveal halos of hot gas. 
This X-ray emitting diffuse gas varies in luminosity 
by a factor of $\sim$100-1000 for a given optical luminosity, or
equivalently, stellar mass (O'Sullivan, Forbes \& Ponman
2001; Fabbiano 2006; Boroson, Kim \& Fabbiano 2011).  It has been suggested that stellar mass loss 
dominates the source of this gas (Sun et al. 2007; Sarzi et al. 2013). However, in a
large X-ray study of ETGs, Goulding et al. (2016) concluded that the data
are not consistent with a simple stellar mass loss picture.

%Ever since this huge scatter in
%hot gas properties was recognised, a large number of works have
%attempted to explain it (see Fabbiano 2006 for a review). 

%A much tighter relation exists between the X-ray luminosity
%(L$_{X,Gas}$) and the temperature of the hot gas (T$_X$), with the gas temperature acting as a 
%proxy for the gravitational potential of the galaxy. 
%The scatter in the relation 
%can be reduced further if restricted to extraction apertures of  
%one effective radius (R$_e$) of the galaxy light (Goulding et al. 2016) or 
%certain galaxy internal properties, such as the presence of a core in the central optical light 
%profile (Kim \& Fabbiano 2015). 

An additional source of gas 
is indicated by the 
cold dark matter (CDM) paradigm for galaxy formation. 
Here infalling pristine gas in massive halos is shock-heated
to X-ray emitting temperatures as these halos collapse. The hot gas
slowly cools while emitting X-rays, with the X-ray luminosity 
directly related to the depth of the potential well (White \& Frenk
1991). In the GIMIC simulations of Crain et al. (2010), the density of
the hot gas is less concentrated than that of the canonical dark
matter density profile. This leads to much lower X-ray luminosities
than predicted by White \& Frenk (1991), which is then closer to 
the X-ray luminosity as observed from late-type galaxy halos. 
However, for
early-type galaxies, AGN heating of hot gas is also potentially
important. 
Recently, Choi et al. (2015) included AGN in 
cosmological models of massive ETGs. They found better agreement 
with the X-ray luminosities of ETGs than when using models 
without AGN feedback, although some discrepancies remained. 
These cosmological simulations all suggest 
that the key parameter determining the luminosity of the diffuse X-ray gas 
(L$_{X,Gas}$) is the galaxy mass (which determines the depth of the potential 
well) with AGN perhaps playing a secondary role in ETGs.

%They argued that the
%hot gas is in quasi-hydrostatic equilibrium, with only minor
%contributions from either inflowing or outflowing winds.  
%McCarthy et al. (2011) 
%incorporated AGN into the OWLS simulation. 
%Although focusing on the group and cluster scales, they 
%showed that AGN are most 
%effective at heating and removing gas at z $\sim$ 2--3  
%(when the galaxy potentials were shallower than today).  

%The presence of an AGN also avoids the traditional
%`overcooling problem' so that the resulting baryon conversion
%efficiencies in these simulations are better matched to results from 
%abundance matching techniques.

%Although AGN are episodic with a duty cycle, they may have
%a large effect on the global X-ray luminosity of galaxy halos.  The
%hot gas in the halo acts as a store of energy from the AGN,
%effectively smoothing out the AGN duty cycle (Kormendy et
%al. 2009; Gasapri et al. 2014). 

Combining new {\it Chandra} data with results from the literature, Kim \&
Fabbiano (2013; KF13) have shown that the dynamical (i.e. baryonic
plus dark matter) mass may indeed be the key parameter in determining
L$_{X,Gas}$ surrounding an ETG 
(see also Mathews et al. 2006). In particular, they showed that
measurements of L$_{X,Gas}$ correlated strongly with the dynamical
mass within 5 R$_e$ (where R$_e$ is the effective radius of the galaxy optical light)
for a sample of 14 early-type galaxies. The total mass measurements
within 5 R$_e$ came from the work of Deason et al. (2012) who used a combination of
planetary nebulae (PNe) and globular cluster (GC) kinematics from the
literature, and the assumption of power-law mass and tracer 
density profiles.

%We note that Su et al. (2015) examined the scatter in the
%global X-ray to K-band luminosity ratio (L$_X$/L$_K$) with the
%enclosed mass within 1 R$_e$ and found no correlation. However, the
%mass within 1 R$_e$ for massive ETGs is dominated by stars; one needs
%to probe further out in galactocentric radius to be dark matter
%dominated. 

%Kim \& Fabbiano (2013) also found an indication that
%galaxies with the %highest masses and X-ray luminosities had the
%tightest relation.

%Recently, Kim \& Fabbiano (2015; KF15) found that slowly rotating
%elliptical galaxies with cores in their optical surface brightness
%profiles showed tighter correlations (e.g. with hot gas temperature),
%but this wasn't extended to the correlation with total mass due to the
%small number of galaxies with dynamical mass measurements available. 

Here, we use a homogeneous analysis of dynamical masses from GC
kinematics within 5 R$_e$ for 29 ETGs (Alabi et al. 2017). This
analysis is based on data from the SLUGGS survey (Brodie et al. 2014)
and supplemented by data from the literature.  X-ray gas luminosities
mostly come from the compilation of Kim \& Fabbiano (2015; hereafter
KF15) and are extracted from radii which vary from 30 to 240 arcsec ($\sim$ 2-5 R$_e$.)
We investigate the trends of the X-ray gas luminosity for our
sample with stellar mass, dynamical mass within 5 R$_e$ and with
extrapolated virial mass. 
We compare these new observations with simulated 
galaxy halos from Crain et al. (2010) and Choi et al. (2015), and
discuss whether the hot gas halos around ETGs are consistent with an 
origin, and heating mechanisms, as described by the $\Lambda$CDM paradigm. 

%So as well as using a fixed scaled radius of 5 R$_e$ %(as per
%Kim \& Fabbiano 2013), we also examine the correlation of %L$_{X,Gas}$
%with a dynamical mass that is matched in radius to that of the X-ray
%emission.

\section{Data}

Globular cluster kinematics for galaxies from the SLUGGS survey 
are supplemented by literature data for NGC 1316, 1399, 4472 and 4636 to
give a total sample of 29 galaxies, 
to derive dynamical masses (Alabi et al. 2017). The
SLUGGS (Brodie et al. 2014) survey targets nearby ETGs 
with stellar masses around 10$^{11}$ M$_{\odot}$. The survey has
accumulated over 4000 high precision GC velocities (see e.g. Pota et
al. 2013). 

%There is a strong overlap with the sample of Deason et
%al. (2012) who used PNe and GCs as dynamical tracers. 

Dynamical masses for the sample galaxies have been derived from the GC
kinematics by Alabi et al. (2017). They used the tracer mass
estimator of Watkins et al. (2010) with the assumption of isotropic
orbits for the GCs (mildly radial or tangential orbits give masses
within $\sim$10\% of the isotropic case). Small corrections have been
made for flattening, rotation or substructures within the GC systems. 
In Fig. 1 we show a comparison
of the total dynamical mass within 5R$_e$ used in this work with that used
by KF13, i.e. from Deason et al. (2012). Although there is a reasonable
match between the two studies, Fig.1 shows that Alabi et al. (2017)
find systematically higher masses for low mass galaxies than Deason et
al. The masses for most of these low mass galaxies were
derived using PNe, which  Alabi et al. (2016) showed 
may underestimate the true galaxy mass compared to other tracers of
mass (i.e. GCs and X-rays). Here we use the Alabi et al. (2017) masses
which are derived only from GCs. Assuming an NFW-like halo, Alabi et
al. (2017) also estimated the total virial mass for each galaxy by
extrapolating from the 5R$_e$ mass.

\begin{figure}
        \includegraphics[angle=-90, width=0.45\textwidth]{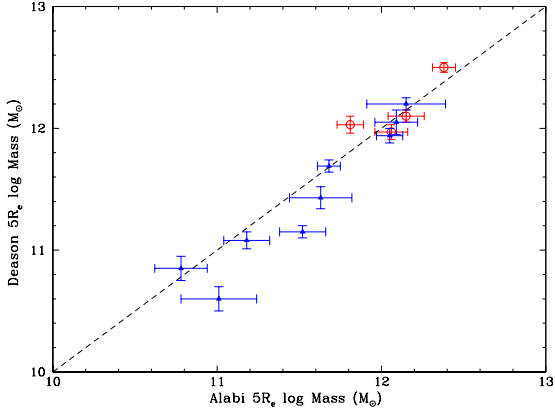}
	    \caption{\label{fig:corrGC2} Comparison of total dynamical mass within 5R$_e$ from 
Deason et al. (2012) with 
that from Alabi et al. (2017) for galaxies used in this work. Deason et al. masses based on globular clusters (GCs) 
are shown 
by open red circles, while those using planetary nebulae (PNe) or a combination of PNe and 
GCs are shown by filled blue triangles. The dashed line shows the 1:1 relation. 
The Alabi et al. masses tend 
to be systematically higher than the Deason et al. masses for lower mass galaxies.  
}

\end{figure} 

%Fragos et al. 2008) 

The diffuse hot gas X-ray luminosities in the 0.3-8 keV band are taken
from the work of KF15. They have carefully removed the contribution
from discrete sources such as low mass X-ray binaries (Fabbiano 2006)
to the total X-ray luminosity, leaving the
diffuse gas contribution L$_{X,Gas}$. A correction to bolometric would increase the X-ray luminosities by 
0.08 dex on average.  Most of the X-ray data come
from {\it Chandra} observations. However, for some high mass galaxies the X-ray emission is
particularly extended (e.g. NGC 4374, 4486, 4649, 5846) and in those
cases {\it ROSAT} data from O`Sullivan, Forbes \& Ponman (2001), corrected
to the {\it Chandra} energy band, are used. Although the contribution from discrete sources in the 
ROSAT data can not be subtracted as accurately as it can for Chandra data, their contribution is only about 
1\% of the diffuse gas luminosity for these high mass galaxies 
(see O'Sullivan, Forbes \& Ponman 2001).  For further details see KF15. 
Here we make a very small correction to the KF15 L$_{X,Gas}$
luminosities for the distances used in the SLUGGS survey (Brodie et
al. 2014).  The KF15 compilation did not include several galaxies that
appear in the Alabi et al. (2017) study. Here we also include the
X-ray luminosities for NGC 720, NGC 1316 and NGC 3115 from Boroson et
al. (2011), and for NGC 5128 from KF13. 
Su et al. (2014) conducted a detailed XMM and {\it Chandra}
study of NGC 1400. As well as the X-ray emission centred on NGC 1400,
they detected an enhanced region of X-rays to the NE of the galaxy
that they associated with stripped gas. Here we use the X-ray
luminosity centred on NGC 1400 with a small adjustment to our X-ray
band and distance, and assume an uncertainty of 20\%. We note that the
X-ray luminosity would double if the enhanced region were also
included. 
Two galaxies in Alabi et al. (2017) but not included here are NGC 2974 (not
observed by {\it Chandra}) and NGC 4474 (the {\it Chandra} observation was only 5
ksec).

In Table 1 we list the properties of our sample galaxies. These 
include the distance-adjusted L$_{X,Gas}$, the presence of a core or
a cusp in the central optical light profile, the local environment of
the galaxy, the stellar mass calculated from the $K$-band luminosity
after the correction described by Scott et al. (2013) 
and  assuming M/L$_K$ = 1,
the dynamical mass within 5 R$_e$ and an estimate of the total virial
mass.

\begin{table*}
\centering
{\small \caption{Galaxy sample and properties}}
\begin{tabular}{@{}lllllllcc}
\hline
\hline
Galaxy & Dist. & R$_e$ & Core & Env.  & log~L$_{X,Gas}$ & log~M$_{\ast}$ & log~M ($<$5~R$_e$) & log~M$_{virial}$\\
$\rm [NGC]$ & [Mpc] 	& ["]  &  &  & [erg/s] & [$\Msun$] &  [$\Msun$] &  [$\Msun$] 	\\
(1) & (2) & (3) & (4) & (5) & (6) & (7) & (8) & (9)\\
\hline
 720 & 26.9 & 35 & 1 & F &  40.68$^\dagger$ & 11.35 & 11.56 (0.20) & 12.97   (0.44)\\
 821 & 23.4 & 40 & 3 & F & 38.40 & 10.97 & 11.63 (0.19) &  12.95    (0.30)\\
1023 & 11.1 & 48 & 3  & G & 38.79 & 10.98 & 11.21  (0.14) &  12.54    (0.28)\\
1316 & 20.8 & 69 & 2 & C & 40.70$^\dagger$ & 11.81 & 12.06 (0.14) & 13.72 (0.44) \\ 
1399 & 20.9 &  56 & 1 & C$^d$ & 41.73 & 11.51 & 12.15  (0.11)   & 13.62    (0.42) \\
1400 & 26.8 & 28 & 1 & G & 39.67$^\mathsection$ & 11.12 & 11.37 (0.23) & 12.71    (0.35)\\
1407 & 26.8 & 63 & 1 & G$^d$ & 41.14 & 11.60 & 12.06  (0.10) & 13.58    (0.44)\\
2768 & 21.8 & 63 & 2 & G & 39.88 & 11.28 & 11.85  (0.13) & 13.23   (0.39)\\
3115 & 9.4 & 35 & 3 & F & 38.38$^\dagger$ & 10.97 & 11.31 (0.14) & 12.63   (0.28)\\
3377 & 10.9 & 36 & 3 & G & 38.00 & 10.44 & 10.78  (0.16) & 12.13    (0.21)\\
3607 & 22.2 & 39 & 1 & G & 40.20 & 11.29 & 11.40  (0.27)  & 12.78    (0.45)\\
3608 & 22.3 & 30 & 1 & G & 39.62 & 10.82 & 11.53  (0.33)  & 12.84    (0.38)\\
%4111 & 14.6 & 12 & -- & 10.66 & &\\
4278 & 15.6 & 32 & 1 & G & 39.39 & 10.88 & 11.45  (0.13)  & 12.76    (0.23)\\ 
4365 & 23.1 & 53 & 1 & G$^d$ & 39.66 & 11.48 & 12.08  (0.10)   & 13.54    (0.42)\\
4374 & 18.5 &  53 & 1 & C & 40.82 & 11.46 & 12.15  (0.24)  & 13.60      (0.46)\\
4459 & 16.0 & 36 & 3 & C & 39.39 & 10.93 & 11.27  (0.33)  & 12.59    (0.41)\\
4472 & 16.7 & 95 & 1 & C & 41.36 & 11.74 & 12.37  (0.10) &  13.95    (0.46)\\
4473 & 15.2 &  27 & 1 & C & 39.09 & 10.87 & 11.16  (0.19) & 12.47    (0.27)\\
4486 & 16.7 &  81 & 1 & C$^d$ & 42.93 & 11.53 & 12.38  (0.07) & 13.87    (0.43)\\
4494 & 16.6 &  49 & 3 & G & 39.11 & 11.02 & 11.18  (0.14) & 12.51    (0.27)\\
4526 & 16.4 &  45 & -- & C & 39.45 & 11.23 & 11.54  (0.18) & 12.90    (0.39)\\
4564 & 15.9 & 20 & 3 & C & 38.59 & 10.58 & 11.01  (0.23) & 12.35    (0.27)\\ 
4636 & 14.3 & 89 & 1 & G$^d$ & 41.50 & 11.13 & 11.81  (0.08) & 13.16   (0.33)\\ 
4649 & 16.5 &  66 & 1 & C & 41.22 & 11.55 & 12.05  (0.08) & 13.55    (0.43)\\ 
4697 & 12.5 & 62 & 3 & G & 39.36 & 11.03 & 11.52  (0.14) &  12.85    (0.27)\\
5128 & 3.8 & 305 & -- & G$^d$ & 40.20$^\ast$ & 10.94 & 11.68 (0.07) & 13.02 (0.23)\\
5846 & 24.2 & 59 & 1 & G$^d$ & 41.70  & 11.40 & 12.09  (0.13) & 13.52    (0.39)\\
5866 & 14.9 & 36 & -- & G & 39.40 & 10.97 & 11.10  (0.40) & 12.42    (0.47)\\
7457 & 12.9 & 36 & 3 & F & 38.08 & 10.28 & 11.05  (0.21) &  12.42    (0.24)\\
\hline
\end{tabular}
\begin{flushleft}
{\small 
Notes: columns are (1) galaxy name, (2) distance, (3) effective radius, (4) 1=core, 2=intermediate, 3=cusp central light profile from Krajnovic et al. (2013) and Lauer et al. (2007), 
(5) Environment: F = field, G = group, C = cluster, $^d$ = central dominant galaxy, (6) 
log X-ray gas luminosity in the 0.3-8 keV band from KF15 (except $\dagger$ = Boroson et al. (2011), $\ast$ = KF13, $\mathsection$ = Su et al. (2014), (7) log total
stellar mass from the $K$-band luminosity (an uncertainty 
of $\pm$0.2 dex allows for reasonable variations in IMF, age and
metallicity), (8) log dynamical mass within 5 R$_e$ and uncertainty, (9) Estimated total virial mass and uncertainty 
(data for columns 7-9 come from Alabi et al. 2017). 
}
\end{flushleft}
\end{table*}

\section{Results and Discussion}

%\begin{figure*}
%        \includegraphics[angle=-90, width=1.05\textwidth]{LX.pdf}
%	    \caption{\label{fig:corrGC2} 
%Diffuse gas X-ray luminosity (0.3-8 keV) vs galaxy mass. Early-type
%galaxies are coded by environment (blue = field, gold = group, green =
%cluster) and central/satellite (triangle = dominant galaxy in large
%group/cluster, circle = satellite). Model galaxies from Choi et
%al. (2015) are shown as stars. {\it Left} panel shows the total
%stellar mass, derived from the K-band luminosity, with an assumed 0.2
%dex uncertainty. The dashed lines show the approximate boundary for
%the galaxy models of Crain et al. (2010) in the 0.5-2 keV X-ray band.
%{\it Middle} panel shows the dynamical mass within a fixed radius of 5
%R$_e$. Some galaxies are highlighted. The solid line shows a weighted
%linear fit to the observations of the form: log L$_{X,Gas}$ =
%3.55($\pm$0.44) Mass ($<$5R$_e$) -- 1.41($\pm$5.15). {\it Right} panel
%shows the estimated total virial mass. The solid and dotted curves
%show the approximate relation and scatter for the galaxy models of
%Crain et al. (2010) in the 0.5-2 keV X-ray band. The models of Crain
%et al. (which focus on late-type galaxies and do not include AGN) need
%to be reduced by a factor of $\sim$50 to match the early-type galaxy
%observations and the models of Choi et al. (which include mechanical
%feedback from AGN). 
%}

%\end{figure*} 

In Fig. 2 we show the X-ray luminosity against stellar mass, 
dynamical mass within 5 R$_e$ (i.e. on a scale that is comparable to the 
X-ray extraction radii used by KF15 and one in which our
sample galaxies are dominated by dark matter) 
and virial mass. These measurements and their uncertainties are given in
Table 1.

We also show in Fig. 2 model predictions from the cosmological
simulations of Crain et al. (2010) and Choi et al. (2015). Crain et
al. (2010) used the GIMIC hydro simulations to predict the X-ray gas
properties of 458 isolated late-type galaxy halos. A key finding of their work
is that the gas is in quasi-hydrostatic equilibrium and that the large
scatter in the optical and IR luminosity--L$_X$ relation is largely
due to the variation of stellar mass at a given halo mass. In contrast, the
halo virial mass--L$_X$ relation is much tighter, since it directly traces
the depth of the potential well. We note that the observations are in
the 0.3-8 keV energy range, while the Crain et al. simulations predict
the 0.5-2 keV range. We expect a small ($<$0.1 dex)
increase in their predicted X-ray luminosities to account for the wider
band in the observations. Crain et al. did not include AGN in their
simulations.

AGN, and the X-ray cavities that they form, are now known to have an
important effect on the hot gas halos around early-type galaxies
(McNamara \& Nulsen 2007; 
Diehl et al. 2008). If an AGN imparts sufficient mechanical
energy to the halo gas it can result in an outflow which reduces the
overall gas density ($n_e$) leading to a significant reduction in the
X-ray luminosity (as L$_{X,Gas} \propto n_e^2$). The Choi et
al. (2015) simulations incorporate  AGN mechanical heating from a wind for a
subsample of 20 ETGs from the cosmological zoom simulations of
Oser et al. (2010). 
Compared to their simulations
without AGN, and AGN with pure thermal heating, AGN with
mechanical heating produce X-ray luminosities that are
100-1000$\times$ lower for a given galaxy.
The simulated galaxies are well matched in stellar mass to that of
the SLUGGS sample, and they predict L$_{X,Gas}$ in the same X-ray
energy band as the observed data. 

%Compared to their simulations
%without AGN, and AGN with pure thermal heating, AGN with
%mechanical heating produce X-ray luminosities that are
%100-1000$\times$ lower for a given galaxy.

%, and in much better
%agreement with observed L$_{X,Gas}$ measurements. %Thus L$_{X,Gas}$
%provides a sensitive test of AGN heating.

In both the Choi et al. and Crain et al. cosmological 
simulations the hot gas is
initially heated to the virial temperature by shocks as it collapses
in the forming dark matter halos. In the case of the Choi et al. simulation, 
feedback from AGN provides additional heating (which reduces the
overall X-ray luminosity since the gas becomes more diffuse due to an
outflow). 

\begin{figure*}
        \includegraphics[angle=-90, width=1.05\textwidth]{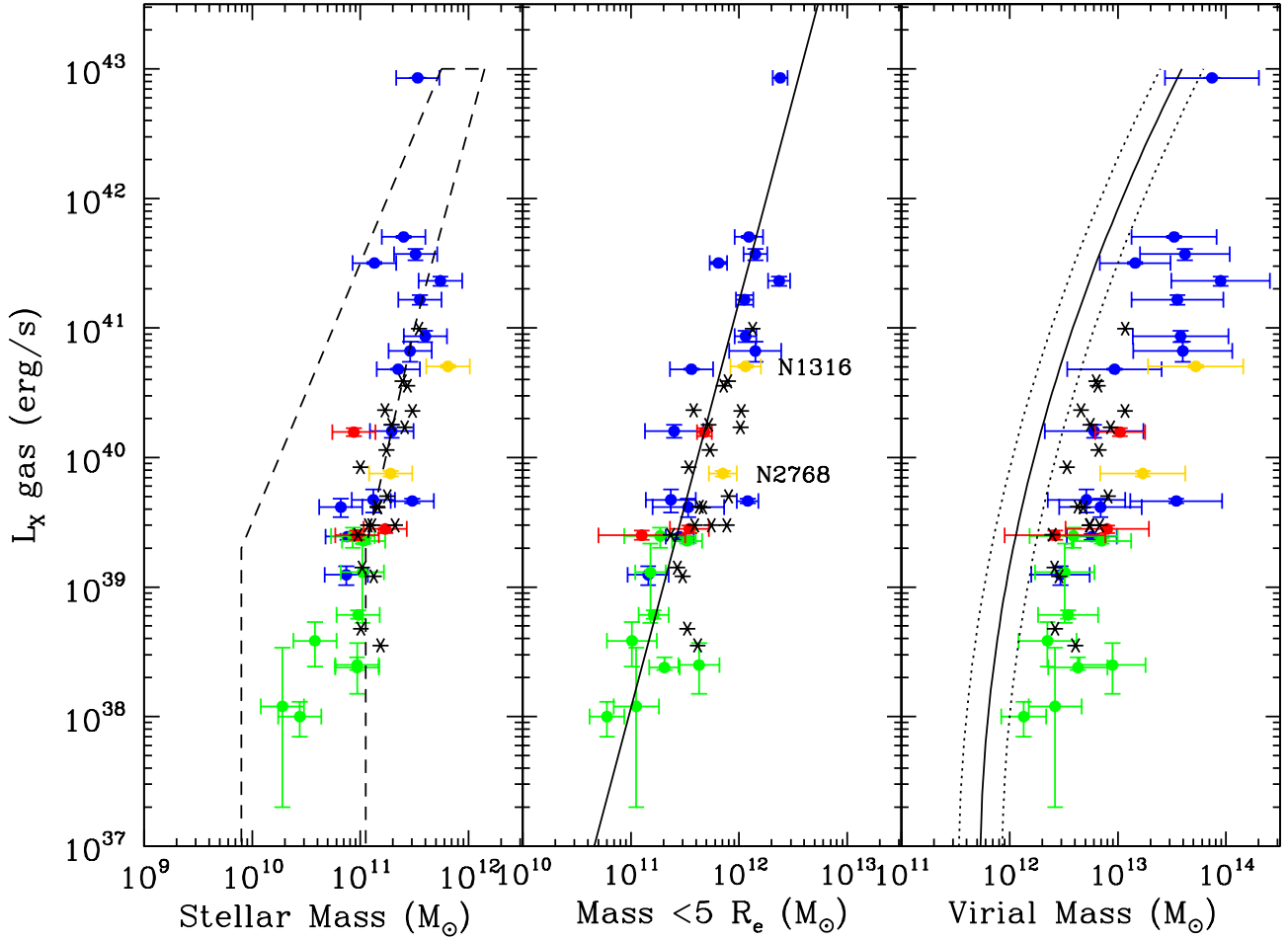}
	    \caption{\label{fig:corrGC2} Diffuse gas X-ray luminosity (0.3-8 keV) vs 
galaxy mass. Early-type galaxies are colour-coded by their central optical light profile (blue = core, gold = intermediate, green = cusp, red = unknown). 
Model galaxies from Choi et al. (2015) are shown as stars. 
{\it Left} panel shows the total stellar mass, derived from the K-band luminosity, with an assumed 0.2 dex uncertainty. 
The dashed lines show the approximate boundary for the galaxy models of Crain et al. (2010) in the 0.5-2 keV X-ray 
band.  
{\it Middle} panel shows the dynamical mass within a fixed radius of 5 R$_e$. The intermediate profile galaxies NGC 2768 and NGC 
1316 are highlighted. The solid line shows a weighted
linear fit to the observations.
% of the form: log L$_{X,Gas}$ =
%3.13($\pm$0.32) log Mass ($<$5R$_e$) +3.64($\pm$3.67).
{\it Right} panel shows the estimated total virial mass. The solid and dotted curves show the approximate 
relation and scatter for the galaxy models of Crain et al. (2010) in the 0.5-2 keV X-ray 
band.  Both core and cuspy galaxies follow a similar trend of X-ray luminosity with galaxy mass. [This figure is best viewed in colour.]
}

\end{figure*}

The left panel of Fig. 2 shows the total stellar mass vs X-ray
luminosity (we note that using the stellar mass within 5R$_e$ would give similar results as 
it is $\sim$90\%, or 0.05 dex, of the total stellar mass for a typical galaxy). The model predictions of Crain et al. (2010) for late-type
galaxies show a large range in X-ray luminosity for a given stellar
mass. The observations for our ETGs reveal a factor of 
100-1000$\times$ range in L$_{X,Gas}$ at a given stellar mass and are
generally located on the right hand edge of the Crain et
al. distribution. On the other hand, the observations are well matched
by the predictions from Choi et al. (2015), albeit over the limited
mass range of their simulations. We suggest that Crain et al. overpredicted 
the X-ray luminosities appropriate for ETGs at a given stellar
mass due to the absence of AGN in their model. We return to this point
below.

%The observations
%also reveal a large range in L$_{X,Gas}$ for a given stellar mass, but
%At a given stellar mass late-type galaxies tend to occupy lower mass 
%halos on average compared to early-type galaxies. Thus the late-type galaxies in 
%the Crain et al. simulation will tend to underpredict the L$_{X,Gas}$ value compared 
%to early-type galaxies.  

In the middle panel we show the dynamical mass within 5 R$_e$ vs X-ray
luminosity. Over this radial range, the galaxies are dark matter
dominated (Alabi et al. 2016). A similar trend was seen by KF13 from
the X-ray emission of 14 ETGs. To determine the best fit we have applied a 
bisector linear regression with errors determined from bootstrap resampling.
This fit is shown in the
middle panel of Fig. 2. This fit is consistent with that found by KF13
(who used masses from Deason et al. 2012), but shifted to slightly
higher masses for low mass galaxies as indicated in Fig. 1. The best fit has a scatter of 0.5 dex about the relation:\\

\noindent
log L$_{X,Gas}$ =
3.13($\pm$0.32) log Mass ($<$5R$_e$) +3.64($\pm$3.67)

We also show the simulated galaxies of Choi et al. (2015) in the
middle panel, and although they only cover a limited range they are
well matched to the observations (Crain et al. did not predict the
mass within 5 R$_e$). The Choi et al. simulations support the
interpretation that the data are well represented by a single linear
relation. We note that Choi et al.  compared their predictions with the
best fit relation of KF13 (which used masses from 
Deason et al. 2013) in their figure 4 and showed that their L$_{X,Gas}$ values are systematically too low 
for a given mass. (When using velocity dispersion as a mass proxy the agreement was very good.
With our larger data
set and the use of (systematically higher) masses from Alabi et
al. (2017) their predictions now agree very well with the
observations of L$_{X,Gas}$ and mass within 5 R$_e$.

%KF13 noted that NGC 4374 (M84) was an outlier in the relation and
%speculated that this was due to ongoing ram pressure stripping of its
%halo within the Virgo cluster (Randall et al. 2008). However, we use a
%lower dynamical mass than Deason et al. (2012) value used by KF13
%(i.e. log M = 12.21) and it now appears to be consistent with the
%steeper trend of the X-ray luminous galaxies.  We also note that NGC
%821 which was another outlier in the KF13 relation continues to be an
%outlier using our dynamical mass. The X-ray luminosity of this galaxy
%can not be affected by ram pressure stripping as it lies in the field
%and is even classified as isolated in some studies (e.g. Reda et
%al. 2004).

In the right panel of Fig. 2 we show the total virial mass vs X-ray
luminosity. The virial mass of the observed ETGs has been estimated
from the dynamical mass measured at 5 R$_e$ and extrapolated to the
virial radius assuming a Navarro, Frenk \& White (1997) 
dark matter profile (see Alabi et al. 2017 for
details). Because the ratio of virial
mass to dynamical mass within 5 R$_e$ 
varies (due to a non-linear stellar mass to halo mass relation), we
expect L$_{X,Gas}$ to reveal a non-linear relation with virial
mass. A non-linear relation can be seen in the observations shown in the 
right panel of Fig. 2.

%Fitting fn is M200 = 0.0485LX^2 - 3.570LX + 77.426

We also show the 20 model galaxies from 
Choi et al. (2015) and the Crain et
al. (2010) predicted trend, and approximate scatter, for their 458 model
galaxies. Crain et al. showed that the scatter in the X-ray luminosity
relation using the virial mass compared to the stellar mass is much
reduced, and argued that the virial mass is the primary driver of the
X-ray luminosity from galaxy halos. Compared to our sample of ETGs,
the Crain et al. model predictions appear to provide an upper envelope;  
their  L$_{X,Gas}$ values
need to be reduced by at least an order of magnitude to match the
observations of ETGs. 
Choi et al. compared 
their predicted L$_{X,Gas}$ values with virial masses for galaxies and groups from Mathews et al. (2006). 
Once again, the Choi et al. (2015) simulations predicted L$_{X,Gas}$ values that are too low 
at a given mass (their figure 4). 
However, with our updated measurements their predictions now reveal a good match to the 
observations of L$_{X,Gas}$ and virial mass. 
%This clearly points to the effects of AGN (included in
%the Choi et al. models but not the Crain et al. ones) in lowering the
%X-ray luminosity. 
%The Crain et al. L$_{X,Gas}$ values
%need to be reduced by at least an order of magnitude to match the
%observations of ETGs. 

We note that several of the galaxies in our sample show `hot cores'
(i.e. central gas temperatures that are hotter than the surrounding
gas) when spatially resolved temperture profiles are available,
e.g. NGC 1316 (Kim \& Fabbiano 2013), 
NGC 3115 (Wong et al. 2011), NGC 4278 (Pellegrini et al. 2012)
and NGC 4649 (Paggi et al. 2014). These hot cores are thought to be due to 
recent heating by the AGN and/or associated radio jets  (e.g. Paggi et al. 2014).
The influence of AGN may extend down to our lowest stellar 
masses of 10$^{10}$ M$_{\odot}$ (Cheung et al. 2016).

KF13 found that the most X-ray
luminous (log L$_{X,Gas}$ $>$ 40) galaxies (i.e. 7 galaxies 
after excluding NGC 4374 which deviated significantly from their trend) 
revealed significantly less scatter than the 6 low X-ray luminosity galaxies.
With our sample of 29 galaxies, and a new 
5 R$_e$ dynamical mass for NGC 4374, we find it lies 
well within the distribution of the X-ray luminous galaxies. We note that 
the $\sim$3 Gyr old merger remnant galaxy NGC 1316 (Fornax A) also lies 
within the scatter of the general trend. 
 
Using a sample of 61 galaxies, KF15 were able to explore 
the L$_{X,Gas}$--T$_X$ relation subdivided by the shape of their 
central optical profile, i.e. core or cusp like. They found that 
core galaxies, which tend to be X-ray luminous, have a very tight relation. 
In contrast, low X-ray luminosity cusp galaxies revealed no clear 
correlation between L$_{X,Gas}$ and T$_X$. 
In Fig. 2 we have colour-coded our sample by their 
central optical profile (see Table 1). 
The galaxies are classified as
core, cusp, intermediate or unknown surface brightness profile
from Krajnovic et al. (2013) or Lauer et al. (2007). We note that there
is some disagreement in the literature about the central
profile in some galaxies, e.g. Dullo \& Graham (2013) suggested that NGC
4473 does not contain a central core but rather a disk.

In all three panels, the sample divides clearly between core 
(high X-ray luminosity) and cusp (low X-ray luminosity)
galaxies, with 5 additional galaxies (3 unknown and 2  
intermediate profiles). 
The sample divides more clearly
in terms of L$_{X,Gas}$ than mass (see also Pellegrini 2005). We note that 
the central profile shape correlates very strongly with central rotation, 
i.e. core galaxies $\approx$ slow rotators, and cusp galaxies $\approx$ fast rotators 
(Krajnovic et al. 2013). 

Focusing on the middle panel, the 17 core galaxies reveal a somewhat
tighter relation than the 9 cuspy galaxies. We find a p-value for the
Spearman rank correlation test of 0.0007 for the core galaxies, whereas the
cusp galaxies have a value of 0.26 indicating almost no relation. A bisector 
fit to only the cuspy galaxies also suggests no statistically 
significant relation.
This supports the findings of KF15 and suggests that SN heating 
may play an increased role in cuspy (low mass) ETGs (David et al. 2006). 
So although the X-ray luminosity of 
both core (high X-ray luminosity) and cuspy (low X-ray luminosity) ETGs are 
primarily determined by total mass, and hence the depth of
the potential well, other factors may play a role for 
the lower mass galaxies.

%It has been suggested that rapid rotation of the galaxy (e.g. a fast central
%rotator) may lead to reduced L$_{X,Gas}$ (Negri et al. 2014). There is
%a strong correlation between the presence of a cuspy profile and fast
%central rotation (Krajnovic et al. 2013). 

%We also examine the relations shown in Fig. 2 coded by environment
%(not shown) but found no compelling trend for cluster galaxies to
%have either reduced L$_{X,Gas}$ due to ram pressure stripping nor
%enhanced L$_{X,Gas}$ due to pressure confinement. A similar conclusion
%was reached by Goulding et al. (2016) in their X-ray analysis.

\section{Conclusions}

Using new dynamical masses we find a strong linear relationship between the diffuse gas X-ray
luminosity for a sample of 29 nearby massive early-type galaxies and their total 
dynamical mass within 5 R$_e$. This result supports the earlier
analysis by KF13 based on 14 galaxies. We also show that the cosmological simulations 
of Choi et al. (2015), which incorporate AGN mechanical heating, 
now agree with observations of the X-ray luminosity and mass within 5 R$_e$.
This good agreement with model galaxies in a cosmological framework
supports the idea that the diffuse X-ray luminosity is primarily
driven by the depth of a galaxy's potential well, with a significant
contribution from AGN mechanical heating. 
Supernova heating may contribute in 
lower mass galaxies. 
%{\bf Given the good agreement with the cosmological simulations of Choi et al., our analysis suggests
%that the majority of the hot gas does not originate from stellar mass
%loss but rather from gas that is shock heated as it falls into a 
%collapsing dark matter halo.}
%Environmental processes, such as ram
%pressure stripping and pressure confinement, appear to have a much smaller 
%role. 

\section{Acknowledgements}

We thank R. Crain for useful discussion and E. Choi for her model data. We also thank the referee for useful 
feedback (of the non-AGN kind).
DAF thanks the ARC for financial support via DP130100388. This work was 
supported by NSF grant AST-1211995 and NASA contract NAS8-03060 (CXC). 
GF thanks the Aspen Center for Physics, funded by NSF grant 1066293, 
for their hospitality while this paper was completed. DWK acknowledges support by the 
Chandra GO grant AR4-15005X and by 2014 Smithsonian Competitive Grant Program for Science.

\section{References}

Alabi A., et al. 2016, MNRAS, in press\\
Alabi A., et al. 2017, in prep\\
Boroson B., Kim D.-W., Fabbiano G., 2011, ApJ, 729, 12 \\
Brodie J.~P., et al., 2014, ApJ, 796, 52 \\
Cheung E., et al., 2016, Natur, 533, 504 \\
Choi E., et al. 2015, MNRAS, 449, 4105 \\
%Choi E., Ostriker J.~P., Naab T., Oser L., Moster B.~P., 2015, MNRAS, 449, 4105 \\
Crain R.~A., et al. 2010, MNRAS, 407, 1403 \\
%Crain R.~A., McCarthy I.~G., Frenk C.~S., Theuns T., Schaye J., 2010, MNRAS, 407, 1403 \\
Diehl S., Li H., Fryer C.~L., Rafferty D., 2008, ApJ, 687, 173-192 \\
Dullo B.~T., Graham A.~W., 2013, ApJ, 768, 36\\
David L.~P., et al. 2006, ApJ, 653, 207 \\
%David L.~P., Jones C., Forman W., Vargas I.~M., Nulsen P., 2006, ApJ, 653, 207 \\
Deason A.~J., et al. 2012, ApJ, 748, 2 \\
%Deason A.~J., Belokurov V., Evans N.~W., McCarthy I.~G., 2012, ApJ, 748, 2 \\
Fabbiano G., 2006, ARA\&A, 44, 323\\ 
Goulding A.~D., et al., 2016, arXiv:1604.01764\\
Kim, D.-W., Fabbiano, G. 2003, ApJ, 586, 826\\
Kim D.-W., Fabbiano G., 2013, ApJ, 776, 116 (KF13)\\
Kim D.-W., Fabbiano G., 2015, ApJ, 812, 127 (KF15)\\
Krajnovi{\'c} D., et al., 2013, MNRAS, 433, 2812 \\
Lauer T.~R., et al., 2007, ApJ, 664, 226\\
Mathews W.~G., et al. 2006, ApJ, 652, L17 \\
%Mathews W.~G., Brighenti F., Faltenbacher A., Buote D.~A., Humphrey P.~J., Gastaldello F., Zappacosta L., 2006, ApJ, 652, L17 \\
McNamara B.~R., Nulsen P.~E.~J., 2007, ARA\&A, 45, 117 \\
Navarro J.~F., Frenk C.~S., White S.~D.~M., 1997, ApJ, 490, 493 \\
Oser L., et al. 2010, ApJ, 725, 2312 \\
%Oser L., Ostriker J.~P., Naab T., Johansson P.~H., Burkert A., 2010, ApJ, 725, 2312 \\
O'Sullivan E., Forbes D.~A., Ponman T.~J., 2001, MNRAS, 328, 461 \\
Paggi A., et al.
%Fabbiano G., Kim D.-W., Pellegrini S., Civano F., Strader J., Luo B., 
2014, ApJ, 787, 134 \\
Pellegrini S., 2005, ApJ, 624, 155 \\
Pellegrini S., et al.  2012, ApJ, 758, 94 \\
%Pellegrini S., Wang J., Fabbiano G., Kim D.-W., Brassington N.~J., Gallagher J.~S., Trinchieri G., Zezas A., 2012, ApJ, 758, 94 \\
Pota V., et al., 2013, MNRAS, 428, 389 \\
Sarzi M., et al., 2013, MNRAS, 432, 1845\\
Scott N., Graham A.~W., Schombert J., 2013, ApJ, 768, 76 \\
Su Y., Gu L., White R.~E., III, Irwin J., 2014, ApJ, 786, 152 \\
Sun M., Donahue M., Voit G.~M., 2007, ApJ, 671, 190\\
Watkins L.~L., Evans N.~W., An J.~H., 2010, MNRAS, 406, 264 \\
White S.~D.~M., Frenk C.~S., 1991, ApJ, 379, 52\\
Wong K.-W., et al. 2011, ApJ, 736, L23 
%Wong K.-W., Irwin J.~A., Yukita M., Million E.~T., Mathews W.~G., Bregman J.~N., 2011, ApJ, 736, L23 \\

\end{document}